%% file: final.tex
\documentclass[letterpaper, 10 pt, conference]{ieeeconf}  

\IEEEoverridecommandlockouts                              
\title{\LARGE \bf
Two-Stage Mechanism Design for Electric Vehicle Charging with Day-Ahead Reservations
}

\author{Pan-Yang Su$^1$, Yi Ju$^2$, Scott Moura$^2$, and Shankar Sastry$^1$
\thanks{This work is supported by Collaborative Research: Transferable, Hierarchical, Expressive, Optimal, Robust, Interpretable NETworks (THEORINET) under Simons Foundation Award No. MPS-MODL-00814647 and NSF Award No. 2031899. We thank Chris Shannon and Manxi Wu for the helpful discussions. We acknowledge the use of ChatGPT and Grammarly for improving the syntax and grammar of several paragraphs.}
\thanks{$^1$EECS, University of California, Berkeley, CA 94720.}
\thanks{$^2$CEE, University of California, Berkeley, CA 94720.}
}

\usepackage{subfigure, cite, graphicx, array, url, hyperref, xstring, algorithm, amsfonts, mathtools, comment, booktabs, multicol, multirow, makecell, xcolor}

\usepackage{amsthm}

\usepackage[noend]{algorithmic}
\let\labelindent\relax
\usepackage[inline]{enumitem}
\usepackage[nameinlink]{cleveref}
\crefname{figure}{Fig.}{Figs.}             
\Crefname{figure}{Figure}{Figures} 
\crefname{table}{Tbl.}{Tbls.}            
\Crefname{table}{Table}{Tables} 
\crefname{algorithm}{Alg.}{Algs.}             
\Crefname{algorithm}{Algorithm}{Algorithms} 
\crefformat{section}{§#2#1#3}
\newcommand{\mkbibdoi}[2][]{doi:~\href{https://doi.org/#2}{\IfStrEq{#1}{}{#2}{#1}}}

\newtheorem{definition}{Definition}[]
\crefname{definition}{Def.}{Defs.}             
\Crefname{definition}{Definition}{Definitions} 

\newtheorem*{alternative design}{Design 1*}
\newtheorem{theorem}{Theorem}[]
\crefname{theorem}{Thm.}{Thms.}             
\Crefname{theorem}{Theorem}{Theorems} 

\newtheorem{proposition}{Proposition}[]
\crefname{proposition}{Prop.}{Props.}             
\Crefname{proposition}{Proposition}{Propositions} 

\newtheorem{assumption}{Assumption}[]
\newtheorem{example}{Example}[]
\newtheorem{remark}{Remark}[]
\crefname{lemma}{Lemma}{Lemmas}             
\Crefname{lemma}{Lemma}{Lemmas} 

\crefname{claim}{Claim}{Claims}             
\Crefname{claim}{Claim}{Claims} 
\begin{document}

\maketitle
\thispagestyle{empty}
\pagestyle{empty}

\begin{abstract}
We analyze the economic (mechanism design) aspect of incorporating flexibility in electric vehicle (EV) charging demand management. We propose a general two-period model, where EVs can reserve charging sessions in the day-ahead market and swap them in the real-time market. Under the model, we explore several candidate mechanisms for running the two markets, compared using several normative properties such as incentive compatibility, efficiency, reservation awareness, and budget balance. Specifically, reservation awareness is the only property coupling the two markets and dictates that an EV will not get a lower utility by joining the real-time market. Focusing on the real-time market, we show that the classical Vickrey-Clarke-Groves (VCG) mechanism that treats the day-ahead allocations as endowments is not budget-balanced. Moreover, we show that no mechanism satisfies some combinations of the properties. Then, we propose using a posted-price mechanism to resolve the issue, which turns out to be the dynamic pricing mechanism adopted in many real-world systems. The proposed mechanism has no efficiency guarantee but satisfies all the other properties. To improve efficiency, we propose using a VCG auction in the day-ahead market that guides the reserve prices in the real-time market. When EVs' valuations in the two markets are close, the proposed approach is approximately efficient. 
\end{abstract}

\section{Introduction}
\subsection{Background}
Pricing electric vehicle (EV) charging for an electric vehicle charging station (EVCS) centers around two main approaches. The first considers a one-shot market such as auctions or competitive equilibria \cite{7548363, 9424480}. While this approach inherits desirable normative guarantees such as efficiency and incentive compatibility from economic theory, it bears an unrealistic assumption that all vehicles arrive at the same time, or at least make decisions simultaneously. Conversely, the second approach assumes that vehicles arrive one at a time and considers online mechanisms \cite{10.5555/2031678.2031733, 7914741} or dynamic pricing \cite{9840998, LEE2020106694, 9875037, 10542458}. It is too pessimistic since some vehicles have known arrival times, so the EVCS can pre-allocate resources before the EVs arrive. Specifically, although some works recognized the uncertainty of future charging demand, they did not allow EVs to reveal estimated valuations or arrival times \cite{10.5555/2031678.2031733, 7914741}.

We propose using ``reservations" to solve the problem. Specifically, we propose a two-period mechanism inspired by the day-ahead and real-time markets in energy systems \cite{1eba3f1b-3634-3b34-9a46-c86b8d069d2d}. EVs in the day-ahead market can reserve charging sessions for the following day. However, they may only have an estimate of the arrival time when making reservations, and the actual realized arrival time may be different from previously expected. To address this issue, we allow EVs to exchange charging sessions in the real-time market. 

Aside from market-based mechanisms, the study of EV charging under a two-stage stochastic optimization framework is similar to our setup \cite{8315146, 6740918, 7394192}. Typically, the operator forms an estimate of EV demand in the day-ahead market and solves a stochastic optimization problem. In the real-time market, actual demand is realized, and the operator needs to update the allocations. Specifically, in this model, EVs can only report their demands in the real-time market. In contrast, we allow EVs to report their demands in the day-ahead market through reservations. This difference leads to two advantages. From the operator's side, she can potentially make better decisions with the additional information. From the EVs' side, they can secure charging sessions in advance, reducing the possibility of arriving at the EVCS without finding an available charging port. However, as previously discussed, ``reservations'' also bring about the complication of co-designing the pricing scheme in both markets, which is absent in previous works where the EVs only participate in the real-time market.

Below, we review some related works that consider a multi-period EV charging scenario with uncertain demand. 

\subsection{Mechanisms without Reservations}
Several previous works considered a multi-period EV charging mechanism from the viewpoint of stochastic optimization involving real-time arriving EV demand uncertainty \cite{6740918}, real-time parking EV flexibility \cite{WU201755, 7360236}, renewable energy sources \cite{7394192, 8293857}, office building management \cite{7833208}, thermal power units \cite{8960365}, multiple microgrids \cite{TAN2022107359}, etc.

However, none of the above works investigates economic aspects. Few works \cite{LIN2023104715, 6808513, 8315146, 10378662} took EVs' disutilities from rescheduling into account, but their considerations are still insufficient. Lin et al. only considered one charging EV \cite{LIN2023104715}, Moeini-Aghtaie et al. failed to account for demand-side uncertainty \cite{6808513}, and Kabiri-Renani et al. restricted the analysis to the day-ahead market without including real-time recourse strategies \cite{10378662}. 

Among these, the model in \cite{8315146} is closest to ours. In  \cite{8315146}, there are two stages, and the central coordinator can reschedule the charging sessions while providing some incentive-based programs to compensate rescheduled EVs. However, there are two key differences. First, their rescheduling refers to the difference between EVs' desired individual schedules and the operator's desired allocation, while our rescheduling accounts for the difference between the realized demand and the estimated demand arising from EVs' inherent uncertainty. Second, while they proposed several incentive programs, they did not offer theoretical analysis. Conversely, we provide a general modeling framework and analyze and propose mechanisms based on well-defined properties.


\subsection{Reservation-Based Mechanisms}
Our work differs from and complements some previous studies that proposed reservation-based mechanisms. On the practical side, Conway demonstrated the effectiveness of the reservation system through simulations \cite{7811294}, Basmadjian et al. proposed an architecture that facilitates the interoperability of different stakeholders \cite{smartcities3040067}, Orcioni and Conti extended the Open Charge Point Protocol standard to include reservations \cite{en13123263}, and Flocea et al. developed and tested a reservation-based EV charging app \cite{s22082834}. However, they focus on testing and implementing algorithms in practice, while we aim to design new mechanisms with theoretical performance guarantees. On the theoretical side, Cao et al. proposed a mechanism that allowed EVs to indicate heterogeneous needs through reservations such as EV types, expected arrival times, and expected charging times, enabling the operator to simultaneously consider parked EVs and remote reservations and better allocate EVs to different EVCSs \cite{8850812}. Following the same architecture, previous works further examined vehicle-to-vehicle (V2V) communication \cite{8734897}, different urgency levels \cite{LIU2021103150}, mobile charging \cite{9068444}, etc. However, none of these works consider uncertainty. 

In our setting, while EVs reveal similar information through reservations, we focus on the interplay between the day-ahead and real-time market of a single EVCS. Specifically, we identify two aspects. First, we consider uncertainty when EVs make reservations and propose a real-time market to reallocate the charging sessions. Second, we treat EVs as self-interested agents and focus on mechanisms that incentivize EVs to report their valuations. 

The only works considering uncertainty in a reservation-based mechanism are \cite{7387372, 7756376, BERNAL2020100388}. However, they either do not consider the market structure and economic incentives \cite{7387372, 7756376} or do not reallocate charging sessions once uncertainty is resolved \cite{BERNAL2020100388}.

\subsection{Proposed Mechanism}
\label{sec: proposed mechanism}
This work is the only one considering demand uncertainty, economic incentives, and multiple periods under a multi-EV scenario with theoretical guarantees; a comparison of related works is given in Appendix \ref{app: compare}. 
However, since we focus on mechanism design, we do not discuss computational issues, and one can easily build the proposed mechanisms on top of existing algorithmic procedures proposed in previous works, e.g., \cite{8315146, 6740918, 7394192}.

The contributions of this work are listed below. All the proofs are relegated to the Appendix.
\begin{enumerate}
    \item We present a general model for two-period (and possibly multi-period) EV charging, cast it as a mechanism design problem, and propose several normative properties to evaluate mechanisms.
    \item We analyze the standard Vickrey–Clarke–Groves (VCG) mechanism \cite[Chapter 9.3.3]{Nisan_Roughgarden_Tardos_Vazirani_2007} and show that it is not directly applicable to our model. Then, we prove several impossibility results, demonstrating that no mechanism features all the normative properties. On the other hand, each mechanism features some subset of the normative properties, thus providing trade-offs across the mechanisms. This paper is the first to detail these trade-offs.
    \item As a practical guide, we propose using a VCG auction in the day-ahead market and a posted-price mechanism in the real-time market. The proposed mechanism is compatible with many existing systems (e.g., \cite{9381520}) and satisfies most of the normative properties.
\end{enumerate}

\begin{remark}
We focus on social welfare instead of revenue for two reasons. First, social welfare is easier to analyze from a mechanism design viewpoint. Second, traditional optimization-based approaches can also be interpreted as maximizing social welfare while treating EVs' cost functions as common knowledge.
\end{remark}



\section{System Model}
We consider a two-period market where the goods are consumed after the second period, both the consumers and the supplier have incentives to buy and sell earlier, and the consumers have uncertain demand in the first period. 

Consider EV charging. A consumer may seek to reserve a charging session so that she can plan her trip in advance. The supplier may wish to sell the charging sessions earlier to optimize system operation. However, there is uncertainty about the arrival time of a vehicle, so, in the first period (\emph{day-ahead market}), the consumer only has an estimate of the arrival times and thus the valuation functions. In the second period (\emph{real-time market}), the consumer may want to change the charging session, so there is a need for re-trade.

\subsection{Formal Definitions}
We model EV charging as the aforementioned two-period market, where the supplier is the EVCS operator, the goods are charging sessions, and the consumers are EVs. Below, we present some definitions, furnished with an example Appendix \ref{app: example}\footnote{Readers may also refer to Examples \ref{ex: bb}-\ref{ex: order}.}. 
Appendix \ref{app: notation} summarizes the notations and includes a flowchart. 

\emph{\textbf{Markets: }}An EVCS operator provides charging service to EVs through an auction mechanism. The mechanism consists of two auctions, occurring at different times (referred to as \emph{periods} or \emph{stages}). The first one, occurring at an earlier time $t = 0$, is called the \emph{day-ahead market}. The second one, occurring at a later time $t = 1$, is called the \emph{real-time market}\footnote{A day-ahead market is also called a forward market, and a real-time market is also called a spot market.}. 


\emph{\textbf{Goods, Bundles, and Allocations:}}
There are $N$ charging ports operating over $T$ \emph{slots}. The set of \emph{goods} is $[T] \times [N]$. The operator provides charging service in the form of charging sessions. A charging session is a series of consecutive slots: $(t, t+1, ..., t + t_d)$ with $t, t_d \in [T], t + t_d \leq T$.  We denote the set of all charging sessions as $\mathbb{T} = \{()\} \cup \{(1), (2), ..., (T)\} \cup \{(1, 2), (2, 3), ..., (T-1, T)\} \cup ... \cup \{(1, 2, ..., T)\}$, which is the set of all the \emph{bundles} in the market. We denote the $j$th bundle in $\mathbb{T}$ as $\mathbb{T}_j$. Note that $\mathbb{T}_1 = ()$ denotes not being allocated. 

\begin{remark}
It is straightforward to introduce more flexibility in defining bundles. For example, to differentiate between different charging powers, one can set the goods as $[T] \times [N] \times [P]$, where $P$ is the number of different charging power levels offered by the operator. 
\end{remark}

An \emph{allocation} is a vector $x \in \{0, 1\}^{|\mathbb{T}|}$ with $\mathbf{1}^T x = 1$, where the $j$th element $x_j = 1$ means that bundle $\mathbb{T}_j$ is allocated and 0 otherwise. 
We also identify an allocation as a vector on the slots, i.e., the operator $\bar{\cdot}$ converts a vector of bundles $x \in \{0, 1\}^{|\mathbb{T}|}$ into a vector of slots $\bar{x} \in \{0, 1\}^{|T|}$ through the following formula.
\begin{equation}
\label{eq: bar x}
\bar{x}_t = \sum_{j \in [|\mathbb{T}|]} x_j \mathbf{I}_{t \in \mathbb{T}_j}, \forall t \in [T].
\end{equation}


\emph{\textbf{Consumers: }}A set $I$ of EVs need to get charged. Each EV $i \in I$ has a valuation vector $v_i \in \mathbb{R}^{|\mathbb{T}|}$, where each element $v_{i, j}$ denotes EV $i$'s monetary valuation of obtaining charging session $\mathbb{T}_j$\footnote{Equivalently, we can identify the valuation function as a mapping from a bundle to a real number, i.e., $v_i: \mathbb{T} \rightarrow \mathbb{R}$.}. In the following, we assume that being unallocated has the least valuation, so after normalization, we have $v_i \in \mathbb{R}_+^{|\mathbb{T}|}$ and $v_{i, 1} = 0$. At time $t = 0$, each EV only has an estimate of $v_i$, described by a cumulative distribution function (CDF) $F_i$ over the set of possible realizations $V_i \subseteq \mathbb{R}_+^{|\mathbb{T}|}$ with full support\footnote{\label{footnote: VNM}We can also endow each EV $i$ with a utility function $u_i$, assumed to be concave and strictly increasing to capture risk aversion. Then, EV $i$'s von Neumann–Morgenstern (VNM) utility function is $U_i = \int u_i(v_i) dF_i \in \mathbb{R}^\mathbb{T}$, and her certainty equivalent is $c_i = u_i^{-1}(U_i)$.}. We denote $V = \Pi_{i\in I}V_i$.
\begin{remark}
$F_i$ can be the subjective probability that EV $i$ ascribes to different valuation vectors and may not correspond to the actual distribution.
\end{remark}

At time $t=1$, uncertainty is resolved, so $v_i$ is perfectly revealed to $i$. For an EV $i \in I$, we use $x_i(0)$ and $x_i(1)$ to denote the allocations in the day-ahead and real-time market, respectively.

\emph{\textbf{Allocation Profile: }}An \emph{allocation profile} is a set of allocations, one for each EV: $x = (x_i)_{i \in I}$\footnote{This notation differs from that in (\ref{eq: bar x}). Here, $x_i \in \{0,1\}^{|\mathbb{T}|}$ represents the allocation of EV $i$, whereas in (\ref{eq: bar x}), $x_j \in \{0,1\}$ denotes the $j$th component of the allocation vector $x \in \{0,1\}^{|\mathbb{T}|}$. Throughout the remainder, $x_i$ will always refer to EV~$i$'s allocation.}. We say an allocation profile is \emph{feasible} if the total allocation for any slot is not greater than the number of charging ports $N$, i.e., $\sum_{i \in I} \bar{x}_i \preceq N\mathbf{1}$. We denote the feasible set as $X$.
\begin{equation}
\label{eq: X}
X = \{x | x_i \in \{0, 1\}^{|\mathbb{T}|}, \mathbf{1}^Tx_i = 1,  \sum_{i \in I} \bar{x}_i \preceq N\mathbf{1}\}.
\end{equation}
\begin{remark}
Throughout this work, we use the shorthand notation $x = (x_i)_{i \in I}$ when considering a profile of elements over $I$. Some examples include $b(0)$, $b(1)$, $p(0)$, $p(1)$, and $p$, as will be seen later.
\end{remark}

Finally, while we do not quantitatively encode the benefits of reservations in our model, it is possible to do so without affecting the theoretical analysis; Appendix \ref{app: reservation} contains more details.  

\subsection{Two-Period Auction Mechanism}
\label{sec: tp model}
If there were no uncertainty and all valuations were common knowledge, the operator could optimally assign EVs to charging sessions. However, valuations are private information known only to the EVs, so the operator needs to devise a \emph{mechanism} and let EVs self-report their valuations, where the reported valuations are referred to as \emph{bids}. However, EVs may not report their valuations truthfully; that is, the bids may differ from the true valuations. For example, an EV may want to overbid (report a bid higher than the valuation) to secure a better charging session. To ensure truthful reporting, the operator additionally charges \emph{payments} from the EVs. For example, an overbidding EV may face a higher payment, resulting in lower overall utility compared to truthful bidding. We formalize these concepts below.

At time 0, EVs submit bids $b(0) = (b_i(0))_{i \in I}$, where $b_i(0) \in \mathbb{R}_+^{|\mathbb{T}|}$ is EV $i$'s \emph{reported} valuation and may differ from $v_i$ due to strategic behavior and uncertainty at time 0. Using $b(0)$, the operator determines an allocation $x(0) = (x_i(0))_{i \in I}$ and payment $p(0) = (p_i(0))_{i \in I}$, where $p_i(0)\in \mathbb{R}$\footnote{We do not require $p_i(0)$ to be non-negative. Under some conditions, $p_i(0) \geq 0$ cannot always hold, as we shall show in later sections. The same holds for $p_i(1)$ defined later.}. 

After that, uncertainty is resolved. 
Then, at time 1, EVs submit bids $b(1) = (b_i(1))_{i \in I}$, where $b_i(1) \in \mathbb{R}_+^{|\mathbb{T}|}$. Using $b(1)$, the operator determines an allocation $x(1) = (x_i(1))_{i \in I}$ and payment $p(1) = (p_i(1))_{i \in I}$, where $p_i(1)\in \mathbb{R}$. Then, the goods are allocated according to $x(1)$ and payments from both periods are collected, so the final allocation profile $x = (x_i)_{i \in I}$ and payment profile $p = (p_i)_{i \in I}$ are 
\begin{equation}
\label{eq: xp}
x_i = x_i(1), p_i = p_i(0) + p_i(1), \forall i \in I.
\end{equation}

Note that the the final allocation $x$ is determined by the real-time allocation $x(1)$, which overrides the day-ahead allocation $x(0)$. In contrast, the final payment $p$ is the sum of the day-ahead payment $p(0)$ and real-time payment $p(1)$, representing the total payment incurred across both periods.

\begin{definition}
\label{def: utility}
Given an allocation $x_i$ and a payment $p_i$, EV $i$'s (quasilinear) utility is $v_i^Tx_i - p_i$.
\end{definition}

We outline the procedure below.
\begin{enumerate}
    \item Each EV $i \in I$ knows $F_i$. The operator knows $V$. 
    \item Each EV $i \in I$ bids $b_i(0)$.
    \item The operator computes an allocation $(x_i(0))_{i \in I}$ and payment $(p_i(0))_{i \in I}$.
    \item Each EV $i \in I$ knows $v_i$ and bids $b_i(1)$.
    \item The operator computes an allocation $(x_i(1))_{i \in I}$ and payment $p_i(1)_{i \in I}$.
    \item The operator implements the allocation $(x_i)_{i \in I}$ and charges the payment $(p_i)_{i \in I}$.
\end{enumerate}

\begin{definition}
\label{def: mechanism}
A \emph{mechanism $M: \mathbb{R}_+^{|\mathbb{T}||I|} \times \mathbb{R}_+^{|\mathbb{T}||I|} \rightarrow X \times \mathbb{R}^{|\mathbb{T}||I|} \times X \times \mathbb{R}^{|\mathbb{T}||I|}$} is a causal mapping from the bids to the allocations and payments.
\begin{equation}
\label{eq: mechanism}
\begin{aligned}
M&: (b(0), b(1)) \mapsto (x(0), p(0), x(1), p(1))\\
&\text{s.t. } b^1(0) = b^2(0) \Rightarrow x^1(0) = x^2(0), p^1(0) = p^2(0).
\end{aligned}
\end{equation}
\end{definition}

The constraint in (\ref{eq: mechanism}) requires that if two bid profiles $b^1$ and $b^2$ have the same day-ahead bids ($b^1(0) = b^2(0)$), the day-ahead allocations and payments should be the same ($x^1(0) = x^2(0), p^1(0) = p^2(0)$). In other words, the day-ahead allocations and payments are independent of future (real-time) bids.


\subsection{Normative Properties}
Ideally, we want the mechanism to satisfy certain desirable (\emph{normative}) properties. The first, \emph{incentive compatibility}, dictates that truthful bidding is a dominant strategy for EVs. This property simplifies EVs' bidding process. 

\begin{definition}
A mechanism is (real-time dominant strategy) incentive-compatible (IC) if truthful bidding ($b_i(1) = v_i$) is a dominant strategy for any uncertainty realization ($v_i \in \arg\max_{b_i(1)} v_i^T x_i(1; b_i(1)) - p_i(b_i(1))$)\footnote{Since $p_i(0)$ is a realized constant in the real-time market, we can replace $v_i \in \arg\max_{b_i(1)} v_i^T x_i(1; b_i(1)) - p_i(1; b_i(1))$ with $v_i \in \arg\max_{b_i(1)} v_i^T x_i(1; b_i(1)) - p_i(b_i(1))$.}.
\end{definition}

\begin{remark}
We use ``$;$'' to indicate the dependence of the variables. For example, $x_i(1; b_i(1))$ means that we focus on the dependence of $x_i(1)$ on $b_i(1)$ while treating the other variables as given, i.e., in the universal quantifier. Thus, $v_i \in \arg\max_{b_i(1)} v_i^T x_i(1; b_i(1)) - p_i(b_i(1))$ is a shorthand for 
\begin{equation}
\begin{aligned}
v_i \in &\arg\max_{b_i(1)} v_i^T x_i(1; b(0), b_i(1), b_{-i}(1)) \\
&- p_i(b(0), b_i(1), b_{-i}(1)), \forall \ b(0), b_{-i}(1).
\end{aligned}
\end{equation}
\end{remark}

Note that we focus on the real-time market in this work; see Appendix \ref{app: remark ic} 
for a more in-depth discussion. Thus, as detailed later in Assumption \ref{assump}, we will impose minimal restrictions on the day-ahead market and treat the outcome $x(0)$ and $p(0)$ as given in the real-time market.

Second, a mechanism should be \emph{individually rational} in the sense that truthful bidding leads to non-negative utility. This property ensures that EVs will be willing to participate in the market.
\begin{definition}
A mechanism is (ex-post) individually rational (IR) if truthful bidding ($b_i(1) = v_i$) leads to non-negative utility ($v_i^Tx_i(1; v_i) - p_i(v_i) \geq 0$).
\end{definition}

Third, for any uncertainty realization, we aim for ex-post efficiency, where the resources are allocated in a way that maximizes social welfare, defined as the summation of all EVs' valuations of the final allocation.
\begin{definition}
A mechanism is (ex-post) efficient (Eff) if $x(1) \in \arg \max_{x(1) \in X} \sum_{i \in I} x_i(1; v)^T v_i$.
\end{definition}

Fourth, the mechanism should be \emph{reservation-aware}. That is, an EV is guaranteed the utility obtained from the day-ahead market $g_i = v_i^Tx_i(0) - p_i(0)$ in the sense that she can always retain the originally allocated good without additional payment if she wants to. Note that reservation awareness is the key feature of the two-period market and implies ex-post individual rationality iff $g_i \geq 0$.
\begin{definition}
A mechanism is reservation-aware (RA) if truthful bidding ($b_i(1) = v_i$) leads to reservation guarantee ($v_i^Tx_i(1; v_i) - p_i(v_i) \geq g_i$).
\end{definition}

Finally, in practice, the operator may not want to pay an EV, so we additionally consider the following two properties. The first property (\emph{budget balance}) dictates that the total payment is non-negative, and the second property (\emph{no subsidy}) requires each individual payment to be non-negative. Note that no subsidy implies weak budget balance. The reason for adopting the stronger notion is to prevent the EV charging market from becoming a financial market where speculators purchase charging sessions in the day-ahead market and profit by reselling them in the real-time market.

\begin{definition}
\label{def: BB}
A mechanism is (weakly) budget-balanced (BB) if the total payment is non-negative ($\sum_{i \in I} p_i(0) + p_i(1) \geq 0$).
\end{definition}

\begin{definition}
A mechanism is no-subsidy (NS) if individual payment is non-negative ($p_i(0) + p_i(1) \geq 0$).
\end{definition}

\begin{remark}
We do not consider revenue maximization in this work. Instead, we focus on other normative properties while making sure the operator does not need to inject money into the market. The discussion of revenue will be future work.
\end{remark}

\subsection{VCG Mechanism}
\label{sec: pre VCG}
We provide some preliminaries of the classical VCG mechanism and establish notations for Section \ref{sec: VCG sec}. We do not specify the time index $t$ since the treatment is one-period. Note that while $X$ can be any general feasible set, throughout this work, it refers to that defined in (\ref{eq: X}).

With a bid profile $b = (b_i)_{i\in I}$ and a feasible set $X$, a VCG mechanism outputs an allocation profile $x^{VCG}(b, X)$ and payment profile $p^{VCG}(b, X)$. The allocation profile $x^{VCG}(b, X)$ is the solution to the social welfare optimization problem (\ref{eq: sw}), whose value is denoted as $SW(b, X)$.
\begin{equation}
\label{eq: sw}
\begin{aligned}
\max_{x \in X} \sum_{i \in I} b_i^T x_i.
\end{aligned}
\end{equation}
With $x^{VCG}(b, X)$, we define the partial sum $SW_{-i}(b, X)$ to be the summation of the social welfare of all the EVs except $i$ under the optimal allocation.
\begin{equation}
SW_{-i}(b, X) = \sum_{j \in I_{-i}} b_j^T x^{VCG}_j(b, X).
\end{equation}

The VCG payment for EV $i$ is the externality she incurs to other agents, defined as the difference in social welfare when she is absent versus when she is present, i.e.,
\begin{equation}
p^{VCG}_i(b, X_{-i}) = SW_{-i}(b_{-i}, X_{-i}) - SW_{-i}(b, X), \forall i \in I.
\end{equation}
Then, $p^{VCG}(b, X) = (p^{VCG}_i(b, X_{-i}))_{i \in I}$. Note that $X_{-i}$ is the set of feasible allocations of all the other EVs when $i$ is absent. Since we ignore day-ahead allocations, $X_{-i}$ is specified by setting $x(0)=\mathbf{0}$ in (\ref{eq: X-i2}).  

One naive approach is to abandon reservations and conduct a single VCG mechanism in the real-time market. We list its property in Table \ref{table: compare}; see \cite[Chapter 9.3]{Nisan_Roughgarden_Tardos_Vazirani_2007} for a proof.


\subsection{Outline}
Table \ref{table: compare} is a comparison of some mechanisms and impossibility results. 
First, in Section \ref{sec: VCG sec}, we will show that conducting a VCG mechanism in the real-time market (TP-VCG) that respects the day-ahead allocations is not budget-balanced. Then, we argue that the negative results are general by showing that incentive compatibility, efficiency, reservation awareness, and budget balance are incompatible in Section \ref{sec: budget balance} and efficiency, reservation awareness, and no subsidy are incompatible in Section \ref{sec: no subsidy}. 

In light of the impossibility results, we propose the notion of constrained efficiency that requires the allocation profile to maximize social welfare under the constraint that no EV $i$ will be allocated a bundle $j$ whose valuation $v_{i, j}$ is less than the utility guarantee $g_i$. Another impossibility result arises. As we show in Section \ref{sec: ce}, any constrained efficient mechanism is not incentive-compatible. Finally, in Section \ref{sec: posted}, we abandon efficiency completely and find out that the posted-price mechanism satisfies all the desirable properties except efficiency. 



\begin{table}
\hspace{0cm}
\caption{A comparison of different mechanisms and some impossibility results (Do not exist: DNE). All cases assume that the mechanism in the day-ahead market is rationalizable as defined in Assumption \ref{assump}.}
\begin{center}
\begin{tabular}{ |c|c|c|c|c|c|c|c| } 
 \hline
 Mechanism & Section & IC & Eff & RA & IR & BB & NS \\ 
 \hline
 VCG & \ref{sec: pre VCG} & \checkmark & \checkmark &  & \checkmark & \checkmark & \checkmark \\ 
 \hline
 TP-VCG & \ref{sec: VCG sec} & \checkmark & \checkmark & \checkmark & & & \\ 
 \hline
 DNE & \ref{sec: budget balance} & \checkmark & \checkmark & \checkmark & & \checkmark & \\ 
 \hline
 DNE & \ref{sec: no subsidy} & & \checkmark & \checkmark &  & & \checkmark \\ 
 \hline
 DNE & \ref{sec: ce} & \checkmark & - & &  & &  \\ 
 \hline
 Posted-Price & \ref{sec: posted} & \checkmark &  & \checkmark & \checkmark & \checkmark & \checkmark \\ 
 \hline
\end{tabular}
\end{center}
\label{table: compare}
\end{table}

\section{Two-Period VCG Mechanism}
\label{sec: VCG sec}
The operator runs an auction by letting EVs submit bids $b(1)$ while treating the day-ahead allocations as EVs' endowments in the real-time market.
\begin{enumerate}[label=(\alph*)]
    \item $b_{i, j}(1) \in \mathbb{R}_+$: Bids should be non-negative.
    \item $b_{i, 1}(1) = 0$: Not being allocated is feasible and has a bid of 0. This ensures the existence of at least one feasible allocation profile.
\end{enumerate}

We set the allocation and payment as follows.
\begin{equation}
x(1) = x^{VCG}(b(1), X), p(1) = (p_i^{VCG}(b(1), X_{-i}))_{i \in I},
\end{equation}
where $X_{-i}$ is given below.
\begin{equation}
\label{eq: X-i2}
\begin{aligned}
&X_{-i} = \{(x_j)_{j \in I_{-i}} | x_j \in \{0, 1\}^{|\mathbb{T}|}, \mathbf{1}^Tx_j = 1,  \\
&\quad\sum_{j \in I_{-i}} \bar{x}_j \preceq N\mathbf{1} - \bar{x}_i(0)\}.
\end{aligned}
\end{equation}

Below, we show the theoretical properties of the two-period VCG (TP-VCG) mechanism. 

%

\begin{proposition}
\label{prop: ic}
The TP-VCG mechanism is incentive-compatible, efficient, and reservation-aware.
\end{proposition}

\section{Three Impossibility Results}
We identify some issues of the above mechanisms and present impossibility results, which bear resemblance to the Myerson–Satterthwaite theorem \cite{MYERSON1983265}; see Appendix \ref{app: MS theorem} 
for a brief review. We assume that any day-ahead allocations and payments rationalizable by a valuation realization is possible.
\begin{definition}
\label{def: rational}
The set of rationalizable day-ahead allocations and payments are
\begin{equation}
\begin{aligned}
E = \{&(x, p) \in X \times \mathbb{R}^{|I|}_+: \exists v \in V \text{ s.t. }\\
&x \in \arg\max_{y\in X}\sum_{i\in I} v_i^T y_i, p_i \in [0, v_i^Tx_i], \forall i \in I \}.
\end{aligned}
\end{equation}
\end{definition}
\begin{assumption}
\label{assump}
We assume that $(x(0), p(0)) \in E$ and any $(x, p) \in E$ is a possible day-ahead outcome.
\end{assumption}

Below, we provide three counterexamples along with more general conditions. Although we do not aim for full generality, the assumptions in Propositions \ref{prop: impos 1}-\ref{prop: impos 3} subsume the following two cases, where $v_{i, j} \in [0, 1], \forall i \in I, j \in \{2, 3, ..., T+1\}$. 
\begin{itemize}
    \item Max-selector: $v_{i, j} = \max_{k \in \mathbb{T}_j} v_{i, k+1}$.
    \item Additive: $v_{i, j} = \sum_{k \in \mathbb{T}_j} v_{i, k+1}$.
\end{itemize}

\subsection{Budget Balance}
\label{sec: budget balance}
While the TP-VCG mechanism ensures efficiency, incentive compatibility, and reservation awareness, it sometimes charges a negative total payment, i.e., the operator subsidizes the EVs. 

\begin{example}
\label{ex: bb}
Consider two EVs and one charging slot allocated to EV 1 in the day-ahead market with a payment of 2, i.e., $x_1 = (0, 1)$ and $p_1(0) = 2$. In the real-time market, EV 1's realized valuation is 7, and EV 2 arrives with a higher valuation of 10. The only efficient allocation is to allocate the charging slot to EV 2, and the TP-VCG payments are $p_1(1) = -10$ and $p_2(1) = 7$, so the total payment is negative, i.e., $\sum_{i\in I} p_i(0) + p_i(1) = -1$.



\end{example}

The problem in Example \ref{ex: bb} is not exclusive to the TP-VCG mechanism, as we show below.

\begin{proposition}
\label{prop: impos 1}
Assume that there exist $U \subseteq V$, $i, j \in I$, $k \in \{2, 3, ..., |\mathbb{T}|\}$, and $x^F \in X$ such that the following hold.
\begin{itemize}
    \item There exists $[a, b]$ with $a < b$ such that for any $v^1, v^2 \in [a, b]$, there exist $v \in U$ with $v_{i, k} = v^1$ and $v_{j, k} = v^2$.
    \item For any $v \in U$ and unique $\{x\} \in \arg\max_{y \in X}v_i^Ty$, we have $[x_{i, k} = 1, x_{j, 1} = 1]$ iff $v_{i, k} \geq v_{j, k}$, $[x_{j, k} = 1, x_{i, 1} = 1]$ iff $v_{j, k} > v_{i, k}$, and $x_\ell = x^F_\ell, \forall \ell \in I \setminus\{i, j\}$.
\end{itemize}
Then, there is no mechanism satisfying incentive compatibility, reservation awareness, efficiency, and budget balance.





\end{proposition}




\subsection{No Subsidy}
\label{sec: no subsidy}
Requiring each individual payment to be non-negative (i.e., no subsidy) leads to a stronger impossibility result: reservation awareness, efficiency, and no subsidy cannot be satisfied simultaneously. 
\begin{example}
\label{ex: incomp}



Consider Example \ref{ex: bb}. The only efficient allocation is to allocate the slot to EV 2. Since $g_1 = 7 - 2 = 5$, to satisfy reservation awareness, the payment of EV 1 must satisfy $p_1(0) + p_1(1) \leq -3$, violating no subsidy.
\end{example}

\begin{proposition}
\label{prop: impos 2}
Assume that there exist $v^1, v^2 \in V$ and $x^1 \in X$ such that $x^1 \in \arg\max_{y\in X} \sum_{i\in I}(v_i^1)^Ty_i$ and $x^1 \notin \arg\max_{y\in X} \sum_{i\in I}(v_i^2)^Ty_i$. Also, there exists $j \in I$ such that $(v^2_j)^T x^1_j > (v^2_j)^Tx^2_j, \forall x^2 \in \arg\max_{y\in X} \sum_{i\in I}(v_i^2)^Ty_i$. Then, there is no mechanism satisfying reservation awareness, efficiency, and no subsidy.
\end{proposition}



\subsection{Constrained Efficiency}
\label{sec: ce}
Example \ref{ex: incomp} demonstrates the conflict between reservation awareness and efficiency. The situation is similar to the school choice problem, where priority and efficiency are generally incompatible \cite{10.1257/000282803322157061}. In our mechanism, we prefer relaxing efficiency; for otherwise, relaxing reservation awareness violates the purpose of incentivizing EVs to participate in the day-ahead market. A minimal requirement for a relaxed version of efficiency is that no EV $i$ should be allocated a bundle which has a lower valuation than the utility guarantee $g_i$. We formalize this notion below.

\begin{definition}
\label{def: ce}
A mechanism is constrained efficient if it maximizes social welfare, with the constraint that $v_i^T x_i(1; v) \geq g_i, \forall i \in I$, i.e., $x \in \arg \max_{x(1) \in X: v_i^T x_i(1; v) \geq g_i, \forall i \in I} \sum_{i \in I} x_i(1; v)^T v_i$.
\end{definition}

\begin{remark}
To gain intuition, suppose $v_i^T x_i(1; v) < g_i$ for some $i \in I$. Then, reservation awareness ($v_i^T x_i(1; v) - p_i \geq g_i$) leads to a negative payment ($p_i < 0$), violating no subsidy. Thus, the constraint rules out such situations.
\end{remark}

Unfortunately, constrained efficiency and incentive compatibility are incompatible.
\begin{example}
\label{ex: ce}


Consider Table \ref{tab: ce}. Suppose there are three EVs ($I = \{1, 2, 3\}$), three slots ($T = 3$), and one charging port ($N = 1$), so the set of goods is $\mathbb{T} = \{(), (1), (2), (3), (1, 2), (2, 3), (1, 2, 3)\}$. EV 1 obtained the first slot with a payment of 2 in the day-ahead market. EVs 2 and 3 got nothing. At the real-time market, EV 2 has a valuation of 13 for the first slot, EV 3 has a valuation of 10 for the second slot, and we consider two situations

When EV 1's realized valuations for the three slots are 7, 7, and 4.5, the constrained efficient outcome is to allocate the second slot to EV 1. When EV 1's realized valuations for the three slots are 7, 9, and 5.5, the constrained efficient outcome is to allocate the third slot to EV 1.

We denote the payment in the first and second situation as $p$ and $q$, respectively. Incentive compatibility implies that $7 - p \geq 4.5 - q$ and $5.5 - q \geq 9 - p$. Then, $5.5 - q \geq 9 - p \geq 6.5 - q$, a contradiction.
\end{example}

\begin{table}
\hspace{0cm}
\begin{center}
\caption{Summary of Example \ref{ex: ce} and Proposition \ref{prop: impos 3}.}
\label{tab: ce}
\begin{tabular}{ |c|c|c|c|c|c| } 
 \hline
 Slot & Day-ahead allocation & $v_1^1$ & $v_1^2$ & $v_2$ & $v_3$ \\ 
 \hline
 1 ($j$) & EV 1 ($p_1(0) = 2$) & 7 & 7 & 13 & 0 \\ 
 \hline
 2 ($k$) &  & 7 & 9 & 0 & 10 \\ 
 \hline
 3 ($\ell$) &  & 4.5 & 5.5 & 0 & 0 \\ 
 \hline
\end{tabular}
\end{center}
\end{table}

\begin{proposition}
\label{prop: impos 3}
Assume that there exist $v^0, v^1, v^2 \in V$, $i \in I$, and $j, k, \ell \in [|\mathbb{T}|]$ such that the following hold.
\begin{itemize}
    \item $v^0_{i, j} = v^1_{i, j} = v^2_{i, j}$ and $v^2_{i, k} - v^1_{i, k} > v^2_{i, \ell} - v^1_{i, \ell}$.
    \item There exists $x^0 \in \arg\max_{y\in X} \sum_{i\in I}(v_i^0)^Ty_i$ such that $x^0_{i, j} = 1$.
    \item For any $x^1 \in \arg\max_{y\in X, y_{i, \ell} = 0} \sum_{i\in I}(v_i^1)^Ty_i$, we have $x^1_{i, k} = 1$ and $(x^1_{i'})^Tv^1_{i'} \geq (x^0_{i'})^Tv^1_{i'}, \forall i' \in I_{-i}$.
    \item For any $x^2 \in \arg\max_{y\in X} \sum_{i\in I}(v_i^2)^Ty_i$, we have $x^1_{i, \ell} = 1$ and $(x^2_{i'})^Tv^2_{i'} \geq (x^0_{i'})^Tv^2_{i'}, \forall i' \in I_{-i}$.
\end{itemize}

Then, there is no mechanism satisfying constrained efficiency and incentive compatibility.
\end{proposition}

\section{Posted-Price Mechanism}
\label{sec: monopoly}
We propose using a posted-price mechanism in the real-time market. A posted-price mechanism is a monopoly economy, which is more applicable to real-world implementations. A monopoly economy can be viewed as imposing strong restrictions on EVs' bids. Specifically, an EV can only bid on one bundle with the reserve price or no bundles. Our model resembles that of \cite{9381520}. 

Specifically, the posted-price mechanism consists of two elements: an EV order and a set of posted prices. The operator queries the EVs based on a predetermined order, and each queried EV selects the desired bundle and pays the corresponding price (or selects the outside option and incurs no payment). In practice, the operator needs not to determine the EVs' order; it can be the EVs' arrival sequence. Thus, the posted-price mechanism is the classical dynamic pricing mechanism.


\subsection{Interpretation}
Aside from the theoretical properties, there are two main reasons for adopting a posted-price mechanism.

\subsubsection{Simplicity}
A posted-price mechanism simplifies EVs' decision-making process. As pointed out in \cite{10.1093/oxrep/grx063}, simple mechanisms encourage participation and reduce cognitive burden, even though such mechanisms limit full preference elicitation.

\subsubsection{Thickness of Real-Time Market}
The real-time market is likely thin, in the sense that few EVs can participate in the auction simultaneously. Thus, even if auctions have good theoretical properties regarding efficiency, they are not practical.

\subsection{A General Posted-Price Mechanism}
\label{sec: posted}
We fix an EV order $\pi: I \rightarrow I$, which is a permutation of $I$. The order $\pi$ is the sequence in which the operator queries the EVs or their order of arrival. We define $\Theta = \{\mathbf{keep}\} \cup [|\mathbb{T}|]$ to be the range of a bid, whose meaning will be defined later. With the EV order $\pi$, we define a set of functions $h_i: \Theta^{i-1} \rightarrow \mathbb{R}_+^{|\mathbb{T}|}$ with $h_{i, 1} = 0$. Each function $h_i$ is a mapping from the previous bids $(b_{\pi(1)}(1), b_{\pi(2)}(1), ..., b_{\pi(i-1)}(1)) \in \Theta^{i-1}$ to a price vector faced by EV $\pi(i)$ while setting the price of not begin allocated as 0. Thus, $h_i$ is the set of reserve prices faced by EV $\pi(i)$.

We denote feasible bundles faced by EV $\pi(i)$ as $\mathbb{T}(i)$. 
Note that we need to update the available bundles after each query since a queried EV may cancel the original allocation.  

\begin{enumerate}
    \item Announce an EV order $\pi: I \rightarrow I$.
    \item Query EVs based on the order $\pi$. Each queried EV $\pi(i)$ faces the available bundles $\mathbb{T}(i)$ and the corresponding prices $h_i((b_{\pi(j)}(1))_{j \in [i-1]})$. 
    Then EV $\pi(i)$ chooses $b_{\pi(i)}(1) \in \Theta$.
    \begin{enumerate}[label=(\alph*)]
    \item If $b_{\pi(i)}(1) = \mathbf{keep}$, set $x_{\pi(i)} = x_i(0)$ and $p_{\pi(i)}(1) = 0$, meaning that she chooses to keep the original allocation without additional payment.
    \item If $b_{\pi(i)}(1) \in [|\mathbb{T}|]$, set $x_{\pi(i)}$ to be an allocation vector with $x_{\pi(i), b_{\pi(i)}(1)} = 1$ and $p_{\pi(i)}(1) = h_{i, b_{\pi(i)}(1)}((b_{\pi(j)}(1))_{j \in [i-1]}) - p_i(0)$, meaning that she cancels the day-ahead allocation with a full refund and re-selects the bundle.
    \end{enumerate}
    \item Update $\mathbb{T}(i+1)$ and compute $h_{i+1}((b_{\pi(j)}(1))_{j \in [i]})$. Go to step 2.
\end{enumerate}

\begin{remark}
While the above procedure is not a simultaneous auction, it still fits into the model in Section \ref{sec: tp model} by invoking the revelation principle \cite[Proposition 9.25]{Nisan_Roughgarden_Tardos_Vazirani_2007}.
\end{remark}


\begin{proposition}
\label{prop: posted}
The posted-price mechanism is incentive-compatible, reservation-aware, individually rational, budget-balanced, and no-subsidy.
\end{proposition}

\subsection{No General Efficiency Guarantee}
Without further assumptions, there is no efficiency guarantee since two EVs may desire each other's bundle, but no order $\pi$ leads to a profitable exchange; see Appendix \ref{app: order} 
for an example.

We do not attempt to fix this type of efficiency loss since it may complicate the mechanism. Instead, we focus on another aspect of efficiency: reserve prices. While historical EV charging data may inform reserve prices, uncertainty is inevitable in the real-time market. To partially resolve uncertainty, we propose conducting a VCG auction in the day-ahead market and use the day-ahead payments to guide the reserve prices in the real-time market.

\subsection{A Simple Reserve Price}
\label{sec: simple reserve}
The posted-price mechanism in Section \ref{sec: posted} is general, and one can instantiate the reserve prices $h_i$'s using different existing algorithms and derive theoretical guarantees under some assumptions. Below, we provide one simple instantiation and derive its efficiency guarantee.

To simplify notation, we relabel EVs such that $\pi(i) = i$. A simple rule is to set the reserve prices equal to the day-ahead payment, regardless of the selected bundle\footnote{For EVs not allocated in the day-ahead market, the reserve prices can be set arbitrarily.}; that is,
\begin{equation}
h_{i, j} = p_i(0), \forall j \neq 0.
\end{equation}

This payment rule has a desirable property: It does not reduce social welfare as long as the realized valuation of the day-ahead allocation is no less than the day-ahead payment ($v_i^T x_i(0) \leq p_i(0)$). In particular, when the day-ahead allocation is approximately efficient with respect to the realized valuations, the resulting real-time allocation will also be approximately efficient.

A more general guarantee is given below, where we define $\delta(v, x(0), p(0))$ to be the summation of those realized valuations which are below the day-ahead payment.
\begin{equation}
\delta(v, x(0), p(0)) = \sum_{i\in I} v_i^T x_i(0) \cdot \mathbf{I}_{v_i^Tx_i(0) < p_i(0)}.
\end{equation}
\begin{proposition}
\label{prop: epsilon efficient}
If the day-ahead allocation is $\epsilon$-efficient ($\sum_{i\in I} v_i^T x_i(0) \geq \max_{y \in X} \sum_{i \in I} v_i^T y_i - \epsilon$) for some $\epsilon \geq 0$, the posted-price mechanism is $(\epsilon + \delta(v, x(0), p(0)))$-efficient ($\sum_{i\in I} v_i^T x_i \geq \max_{y \in X} \sum_{i \in I} v_i^T y_i - \epsilon - \delta(v, x(0), p(0))$).
\end{proposition}

While the posted-price mechanism can be applied with any day-ahead allocation and payment rules, using a VCG mechanism in the day-ahead market is a sensible option, as it offers strong theoretical guarantees (approximate efficiency and incentive compatibility) when uncertainty is minimal. Further details are provided in Appendix \ref{app: day-ahead vcg}. 

\section{Conclusions}
We proposed a two-stage EV charging mechanism that accounts for agents' uncertainty of valuations (e.g., arrival or departure times) and the desire to reserve charging sessions beforehand. We showed that no mechanism satisfies all the desirable properties and proved that a real-time posted-price mechanism along with an approximately efficient day-ahead auction satisfies most properties and aligns well with practical implementations. 

For future work, we will conduct simulations on real-world data to gauge the significance of the impossibility results, which are of a worst-case flavor, and we aim to derive more general theories and apply the model to other applications involving multiple periods. Moreover, while we advocate for combining an approximately efficient day-ahead auction with a real-time posted-price mechanism, there remains substantial flexibility in the specific design choices. Exploring concrete instantiations under different practical assumptions or suitably relaxed normative properties is an interesting direction for future research.

\bibliographystyle{IEEEtran}
\bibliography{reference}



\input{Appendix_final}

\end{document}

%% file: Appendix_final.tex
\clearpage
\appendix
\subsection{Comparison of Models}
\label{app: compare}
We compare the system models from three aspects in Table \ref{table: related work}. Note that while some related works \cite{10.5555/2031678.2031733, 7914741, 9840998, LEE2020106694, 9875037, 10542458} recognized the existence of uncertain demand, the operator makes decisions after uncertainty is resolved, i.e., after EVs arrive at the EVCS.

\begin{table}[h]
\hspace{0cm}
\caption{A comparison of the models in previous works and the proposed one.}
\begin{center}
\begin{tabular}{ |c|c|c|c| } 
 \hline
 Model & \thead{Demand\\ uncertainty} & \thead{Economic\\ incentives} & \thead{Multiple\\ periods} \\ 
 \hline
 \cite{7548363, 9424480} &  & \checkmark & \\
 \hline 
 \cite{10.5555/2031678.2031733, 7914741, 9840998, LEE2020106694, 9875037, 10542458} & - & \checkmark & \\
 \hline
 \cite{6740918, WU201755, 7360236, 7394192, 8293857, 7833208, 8960365, TAN2022107359, 7387372, 7756376} & \checkmark & & \checkmark \\
 \hline
 \cite{6808513} & & \checkmark & \checkmark \\
 \hline
 \cite{7811294,smartcities3040067,en13123263,s22082834,8850812,8734897,LIU2021103150,9068444} & & & \checkmark\\
 \hline
 \cite{10378662, BERNAL2020100388} & \checkmark & \checkmark & \\
 \hline
 \cite{8315146, LIN2023104715}, Proposed & \checkmark & \checkmark & \checkmark \\
 \hline
\end{tabular}
\end{center}
\label{table: related work}
\end{table}

\subsection{An Example of Real-Time Market}
\label{app: example}
\begin{example}
\label{ex: example}
Suppose there are two EVs ($I = \{1, 2\}$), two slots ($T = 2$), and one charging port ($N = 1$), so the set of goods is $\mathbb{T} = \{(), (1), (2), (1, 2)\}$. In the real-time market, EV 1 only needs one slot to fully charge her EV and prefers charging earlier. Specifically, she has a valuation of each slot for 2 and 7, respectively, but getting both slots does not increase her valuation ($v_1 = (0, 7, 2, 7)$). EV 2 also needs one slot, but cannot arrive early and is willing to pay more to charge her EV ($v_2 = (0, 0, 10, 10)$).

To maximize social welfare, the operator allocates the first slot to EV 1 and the second slot to EV 2, so EV 1 gets $\mathbb{T}_2$ and EV 2 gets $\mathbb{T}_3$ ($x_1(1) = (0, 1, 0, 0)$ and $x_2(1) = (0, 0, 1, 0)$). 

We summarize the example below.
    \begin{itemize}
        \item $T = 2$, $N = 1$, and $\mathbb{T} = \{(), (1), (2), (1, 2)\}$. 
        \item $I = \{1, 2\}$, $v_1 = (0, 7, 2, 7)$, and $v_2 =(0, 0, 10, 10)$.
        \item $x_1(1) = (0, 1, 0, 0)$ and $x_2(1) = (0, 0, 1, 0)$.
        \item $\bar{x}_1(1) = (1, 0)$ and $\bar{x}_2(1) = (0, 1)$.
\begin{table}[h]
\caption{Summary of Example \ref{ex: example}.}
\begin{center}
\begin{tabular}{ |c|c|c|c| } 
 \hline
 Slot & Real-time allocation & $v_1$ & $v_2$ \\ 
 \hline
 1 & EV 1 & 7 & 0 \\ 
 \hline
 2 & EV 2 & 2 & 10 \\ 
 \hline
\end{tabular}
\end{center}
\end{table}
\end{itemize}
\end{example}

\subsection{Notation and a Flowchart}
\begin{figure*}
    \centering
    \includegraphics[width=.8\linewidth]{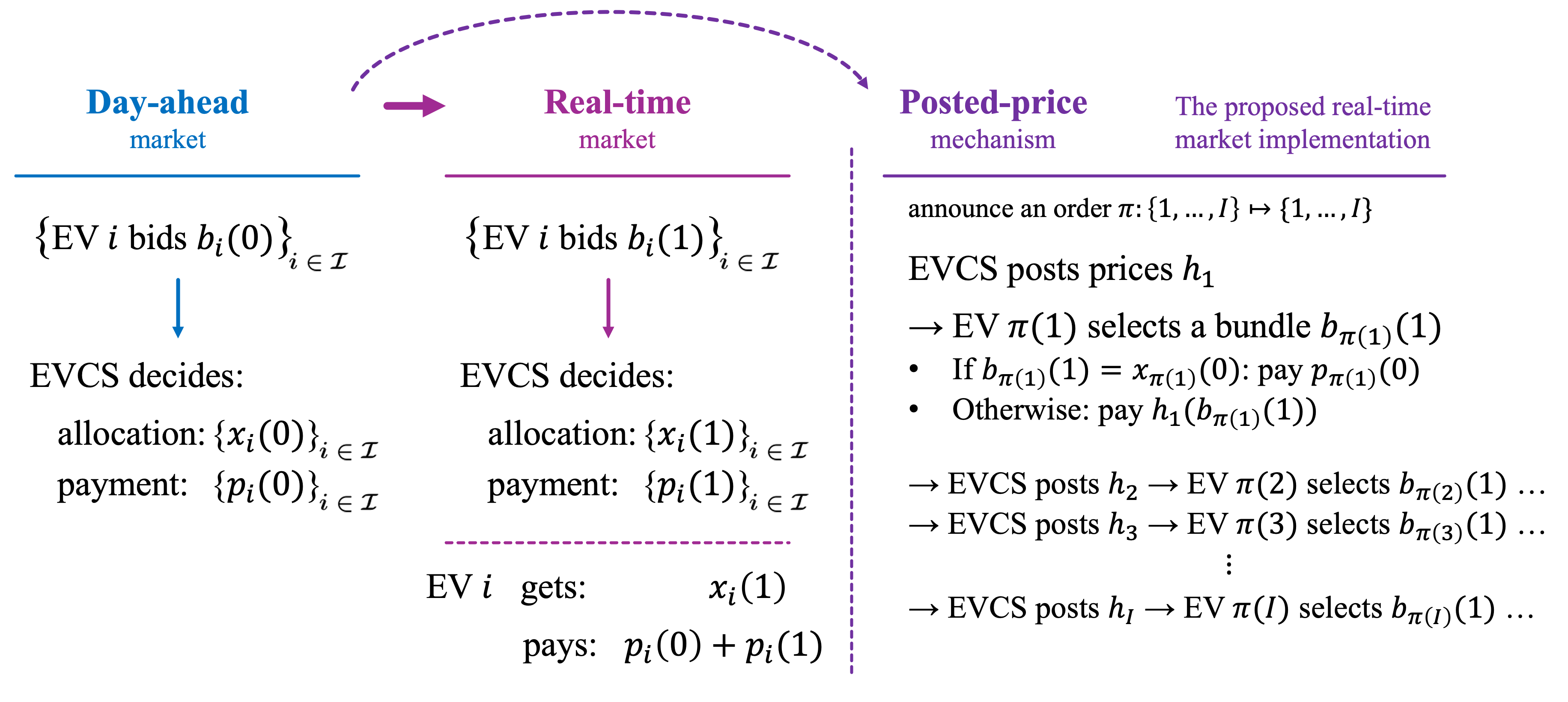}
    \caption{Flowchart of the two-period auction mechanism.}
    \label{fig:flowchart}
\end{figure*}

Table \ref{table: notation} summarizes the notations in this work. Note that in (\ref{eq: bar x}) and the surrounding discussion, the symbol $x$ denotes a single allocation, while in all other places, it refers to an allocation profile. Additional mathematical notations are listed below, and Fig. \ref{fig:flowchart} is a flowchart of the two-period auction mechanism.

\emph{Notation: }We denote the set of real numbers by \(\mathbb{R}\), non-negative real numbers by \(\mathbb{R}_+\), integers by \(\mathbb{Z}\), non-negative integers by \(\mathbb{Z}_+\), and natural numbers by \(\mathbb{N}\). For $N \in \mathbb{N}$, we define $[N] := \{1, 2, ..., N\}$. When constructing vectors, we use the following notation: $b = (b_1, b_2, ..., b_N) = [b_1 \ b_2 \ ... \ b_N]^T$.  We denote a 1-vector by $\mathbf{1} := (1, 1, ..., 1)$ and a 0-vector by $\mathbf{0} := (0, 0, ..., 0)$, with dimensions clear from the context. We denote the indicator function by $\mathbf{I}_{(\cdot)}$, which is 1 when $(\cdot)$ is true and 0 otherwise. When indexing a vector $b = (b_1, b_2, ..., b_N)$, we follow the standard game-theoretic notation: $b_{-i} := (b_1, ..., b_{i-1}, b_{i+1}, ..., b_N)$ and $b = (b_i, b_{-i})$. Sometimes, we treat a set as both a set and a vector to simplify notation, i.e., $b = \{b_1, b_2, ..., b_N\} = (b_1, b_2, ..., b_N)$.

\label{app: notation}
\begin{table}[h]
\begin{center}
\caption{Table of Notations}
\label{tab:Table of Notations}    
\begin{tabular}[t]{p{1.8cm}p{5.4cm}}
\hline\hline
{\bf Notation} & {\bf Definition} \\
\hline\hline
$N$ & Number of charging ports \\
$T$ & Number of time slots \\
$[T] \times [N]$ & Set of goods \\
$I$ & Set of EVs \\
$(t, ..., t+t_d)$ & A charging session with start time $t$ and end time $t+t_d$ \\
$\mathbb{T}$ & Set of all the bundles \\
& $\mathbb{T} = \{()\} \cup \{(1), (2), ..., (T)\} \cup \{(1, 2), (2, 3), ..., (T-1, T)\} \cup ... \cup \{(1, 2, ..., T)\}$ \\
$\mathbb{T}_j$ & The $j$th bundle in $\mathbb{T}$ ($\mathbb{T}_1 = ()$ denotes not being allocated) \\
$x$ & Allocation ($x_j = 1$ denotes that bundle $\mathbb{T}_j$ is allocated) \\
$\bar{x}$ & Converted vector of bundles (defined in (\ref{eq: bar x})) \\
$X$ & Feasible set of allocations (defined in (\ref{eq: X})) \\
\hline
$v_{i, j}$ & EV $i$'s monetary valuation of obtaining charging session $\mathbb{T}_j$ \\
$v_i$ & EV $i$'s valuation vector \\
$V_i$ & Set of possible realizations of $v_i$ \\
$F_i$ & EV $i$'s estimated CDF of $v_i$ in the day-ahead market (with support being $V_i$) \\
$V$ & Support profile defined by $V = \Pi_{i \in I} V_i$ \\
\hline
$b_i(0)$ / $b_i(1)$ & EV $i$'s bid in the day-ahead/real-time market \\
$b(0)$ / $b(1)$ & Bid profile in the day-ahead/real-time market\\
& $b(0) = (b_i(0))_{i \in I}$ / $b(1) = (b_i(1))_{i \in I}$ \\
$x_i(0)$ / $x_i(1)$ & EV $i$'s allocation in the day-ahead/real-time market \\
$x(0)$ / $x(1)$ & Allocation profile in the day-ahead/real-time market \\
& $x(0) = (x_i(0))_{i \in I}$ / $x(1) = (x_i(1))_{i \in I}$ \\
$p_i(0)$ / $p_i(1)$ & EV $i$'s payment in the day-ahead/real-time market \\
$p(0)$ / $p(1)$ & Payment profile in the day-ahead/real-time market \\
& $p(0) = (p_i(0))_{i \in I}$ / $p(1) = (p_i(1))_{i \in I}$ \\
\hline
$x_i$ / $p_i$ & EV $i$'s final allocation/payment \\
& $x_i = x_i(1)$ / $p_i = p_i(0) + p_i(1)$ \\
$x$ / $p$ & Final allocation/payment profile \\
& $x = (x_i)_{i \in I}$ / $p = (p_i)_{i \in I}$ \\
EV $i$'s utility & $v_i^Tx_i - p_i$ (Definition \ref{def: utility}) \\
$M$ & Mechanism (Definition \ref{def: mechanism}) \\
$E$ & Set of rationalizable day-ahead allocations and payments (Definition \ref{def: rational}) \\
\hline
\end{tabular}
\label{table: notation}
\end{center}
\end{table}

\subsection{Benefits of Reservations}
\label{app: reservation}
We can quantitatively encode the benefits of reservations in our model. 

From the EVs' viewpoint, not being allocated in the day-ahead market and not being allocated in the real-time market are different. In the former case, the EV has zero valuation. In the latter case, the EV may have a negative valuation since she wastes the time of driving to the charging session without being charged.

From the operator's viewpoint, her operating cost critically depends on her time-of-use energy cost and demand charge. Specifically, spreading out the demand is beneficial for decreasing the operating cost. Therefore, if the operator has an estimate of the demand, perhaps obtained from the day-ahead bids, she can set a higher real-time price at the demand peak to reduce the demand charge.

As long as these benefits or costs are exogenously given, we can include them in our model without affecting the theoretical analysis.

\subsection{A Focus on Real-Time Market}
\label{app: remark ic}
With the notation in footnote \ref{footnote: VNM}, we can also define other notions of incentive compatibility similarly. For example, a mechanism is ex-post incentive-compatible if $b_i(0) = U_i$ and $b_i(1) = v_i$ is an ex-post dominant strategy for any uncertainty realization and is ex-ante incentive-compatible if $b_i(0) = U_i$ and $b_i(1) = v_i$ is an ex-ante dominant strategy. However, in general, when there is uncertainty, ex-post incentive compatibility is unattainable, and ex-ante incentive compatibility is difficult to define\footnote{For example, when the valuation of a charging session $v_{i, j}$ is a uniform distribution on $[10, 20]$, any bid $b_i \in (10, 20)$ can be justified by a suitably constructed utility function $u_i$. Thus, any bid $b_i \in (10, 20)$ can be considered truthful.}. Therefore, we focus on the well-defined and achievable notion of real-time incentive compatibility.

Specifically, the valuation profile is a probability distribution, so the bidding strategy depends on the risk aversion level of the consumer. Thus, while one can define incentive compatibility and other properties in terms of the certainty equivalent ($b_i(0) = c_i$) in the day-ahead market, we refrain from fully characterizing the equilibrium under the mechanism as it requires players to behave sophisticatedly. Instead, we aim to derive general results and will focus on analyzing the real-time market.

\subsection{Myerson-Satterthwaite Theorem}
\label{app: MS theorem}
We state \cite[Corollary 1]{MYERSON1983265} without attempting a rigorous treatment. For our purposes, we consider a more restricted setting in which the desired properties hold ex post. This suffices for the results developed in this work. The original version is stated in terms of expectations and could be used to derive a more general version of our results.

In the bilateral trading problem, the objective is to devise a mechanism (allocation and payment) for trading a good between a seller and a buyer. The mechanism solicits reported valuations (bids) from both players, and decide the probability for the trade to occur, the amount the buyer has to pay, and the amount the seller can receive. Note that Myerson and Satterthwaite define a mechanism as one without receiving outside money, so it is by definition budget balanced according to our Definition \ref{def: BB}.

\begin{theorem}[Myerson-Satterthwaite Theorem \cite{MYERSON1983265}]
If the seller’s valuation is distributed with positive
probability density over the interval $[a_1, b_1]$, and the buyer’s valuation is distributed with positive probability density over the interval $[a_2, b_2]$, and if the interiors of these intervals have a nonempty intersection, then no incentive-compatible individually rational trading mechanism can be ex post efficient.
\end{theorem}

\subsection{Proof of Proposition \ref{prop: ic}}
\label{app: proof ic}
\begin{proof}
\begin{enumerate}
    \item \emph{Incentive Compatibility:} We follow the standard argument. With the VCG payment, EV $i$'s utility when bidding $b_i(1) = v_i$ is
\begin{equation}
\begin{aligned}
&v_i^T x^{VCG}_i(b(1), X) + SW_{-i}(b(1), X) \\
&\quad- SW_{-i}(b_{-i}(1), X_{-i}) - p_i(0).
\end{aligned}
\end{equation}
The last two terms are independent of $b_i(1)$, so EV $i$ will choose $b_i(1)$ that maximizes $v_i^T x^{VCG}_i(b(1), X) + SW_{-i}(b(1), X)$, which is the social welfare at $b_i(1) = v_i$.

Consider any other bid $b_i'(1) \neq b_i(1)$. Social welfare at $b_i'(1)$ is
\begin{equation}
v_i^T x^{VCG}_i(b'(1), X) + SW_{-i}(b'(1), X).
\end{equation}


Since $x^{VCG}(b(1), X)$ maximizes social welfare, we have $v_i^T x^{VCG}_i(b(1), X) + SW_{-i}(b(1), X) \geq v_i^T x^{VCG}_i(b'(1), X) + SW_{-i}(b'(1), X)$.

    \item \emph{Efficienty:} This follows directly from the optimization problem (\ref{eq: sw}).

    \item \emph{Reservation Awareness:} The mechanism maximizes social welfare. In particular, social welfare under $x^{VCG}(b(1), X)$ is at least that under $x'$ with $x'_i = x_i(0)$ and $x'_{-i} = x^{VCG}(b_{-i}(1), X_{-i})$.
\begin{equation}
\begin{aligned}
&v_i^T x^{VCG}_i(b(1), X) + SW_{-i}(b(1), X) \\
&\geq v_i^Tx_i(0) + SW_{-i}(b_{-i}(1), X_{-i}).
\end{aligned}
\end{equation}

Rearranging terms gives the result.
\begin{equation}
\begin{aligned}
&v_i^T x^{VCG}_i(b(1), X) + SW_{-i}(b(1), X) \\
&\quad - SW_{-i}(b_{-i}(1), X_{-i}) - p_i(0)\\
&\geq  v_i^Tx_i(0) - p_i(0).
\end{aligned}
\end{equation}
\end{enumerate}
\end{proof}

\subsection{Proof of Proposition \ref{prop: impos 1}}
\label{app: impos 1}
\begin{itemize}
    \item Proof: 
    
    We provide some intuition before the formal proof. The first condition ensures that the supports of EV $i$'s and EV $j$'s valuations overlap. The second condition guarantees that there is a set of valuations such that bundle $k$ is only given to EV $i$ or EV $j$, but not any other EV. Together, these conditions isolate a two-agent (EV $i$ and EV $j$) setting over a single good (bundle $k$), enabling the application of the Myerson–Satterthwaite theorem to establish the impossibility result.
    
    Formally, for any feasible $x$, since $(x, \mathbf{0})$ is rationalizable, we must have $p_\ell(1) \leq 0, \forall \ell \in I \setminus\{i, j\}$ to satisfy reservation awareness. 
Thus, the total payment is $\sum_{\ell\in I} p_\ell \leq p_i + p_j$, and the result follows from the Myerson–Satterthwaite theorem \cite{MYERSON1983265}.



    \item The proposition is sharp. Consider Example \ref{ex: bb} and relax any of the four properties. The TP-VCG mechanism does not satisfy budget balance. The mechanism that allocates the slot to EV 2 and has her pay EV 1 seven dollars is not incentive-compatible, as EV 1 can overbid to eight and get a higher payment. A mechanism that follows the day-ahead allocation and payment is not efficient, and the VCG mechanism in Subsection \ref{sec: pre VCG} fails to satisfy reservation awareness.


\end{itemize}

\subsection{Proof of Proposition \ref{prop: impos 2}}
\label{app: impos 2}
\begin{itemize}
    \item Proof: 
    
    With the stated assumption, it is possible to have a day-ahead allocation $x^1$ (that is rationlizable based on the valuation profile $v^1$), but the realized real-time valuation profile turns out to be  $v^2$. In such a case, there is an EV $j$ that cannot be reallocated in the real-time market in a way that satisfies the reservation guarantee.
    
    Formally, when $x(0) = x^1$, $p_j(0) = 0$, and $v = v^2$, we have $g_j = (x^1_j)^Tv^2_j$. Fix any efficient real-time allocation $x^2$ and EV $j$'s payment $p_j$. Since $(v^2_j)^T x^1_j > (v^2_j)^Tx^2_j$, we have $(v^2_j)^Tx^2_j - p_j < (v^2_j)^Tx^1_j = g_j$ for any $p_j \geq 0$. Thus, no efficient real-time allocation and non-negative payment satisfy reservation awareness.
    \item Proposition \ref{prop: impos 2} is sharp in that there exist mechanisms satisfying any two properties. The TP-VCG mechanism satisfies reservation awareness and efficiency. The modified mechanism based on Definition \ref{def: ce} satisfies reservation awareness and no subsidy. The VCG mechanism without treating $x(0)$ as endowment in the real-time market satisfies efficiency and no subsidy.
\end{itemize}

\subsection{Proof of Proposition \ref{prop: impos 3}}
\label{app: impos 3}
\begin{itemize}
    \item Proof: With the stated assumption, there exists some rationalizable $(x^0, p^0)$ such that $x^0_{i, j} = 1$ and $v^2_{i, j} - v^2_{i, \ell}< p^0_i < v^1_{i, j} - v^1_{i, \ell}$. Under $v^1$, any constrained efficient allocation $x^1$ must have $x^1_{i, \ell} = 0$, as $v^1_{i, \ell}  < v^1_{i, j} - p^0_i$. Thus, any constrained efficient outcome $x^1$ must have $x^1_{i, k} = 1$. Under $v^2$, any constrained efficient outcome $x^2$ must have $x^2_{i, \ell} = 1$.

We denote the payment under $v^1$ and $v^2$ as $p$ and $q$ respectively, and incentive compatibility implies the following.
\begin{equation}
v^1_{i, k} - p \geq v^1_{i, \ell} - q
\end{equation}
\begin{equation}
v^2_{i, \ell} - q \geq v^2_{i, k} - p
\end{equation}
Then, $v^2_{i, \ell} + v^1_{i, k} - v^2_{i, k} - q \geq v^1_{i, k} - p \geq v^1_{i, \ell} - q$, which contradicts $v^2_{i, k} - v^1_{i, k} > v^2_{i, \ell} - v^1_{i, \ell}$.
    \item The reason underlying the impossibility result is that the allocation function does not satisfy weak monotonicity \cite{https://doi.org/10.1111/j.1468-0262.2006.00695.x}.
\end{itemize}

\subsection{Proof of Proposition \ref{prop: posted}}
\label{app: posted}
\begin{proof}
Incentive compatibility is immediate since every EV $i$ only needs to select the bundle with the highest utility. Reservation awareness and individual rationality then follow by observing that EV $i$ has the option to keep the original bundle without incurring additional payment or choosing nothing while receiving a full refund.

Budget balance and no subsidy are immediate since all the prices are non-negative.
\end{proof}

\subsection{An Example of Posted-Price Mechanism}
\label{app: order}
\begin{example}
\label{ex: order}
Suppose there are two EVs ($I = \{1, 2\}$), two slots ($T = 2$), and one charging port ($N = 1$), so the set of goods is $\mathbb{T} = \{(), (1), (2), (1, 2)\}$. EV 1 obtained the first slot with a payment of 1 in the day-ahead market ($x_1(0) = (0, 1, 0, 0)$ and $p_1(0) = 1$). EV 2 obtained the second slot with a payment of 1 in the day-ahead market ($x_2(0) = (0, 0, 1, 0)$ and $p_2(0) = 1$). In the real-time market, EV 1 has a valuation of each slot for 2 and 7, respectively ($v_2 = (0, 2, 7, 7)$), and EV 2 has a valuation of 7 and 2 ($v_2 = (0, 7, 2, 7)$).

The only efficient allocation is to let EVs exchange charging slots, but every EV order $\pi$ ($\pi(1) = 1, \pi(2) = 2$ and $\pi(1) = 2, \pi(2) = 1$) will result in EVs keeping their day-ahead allocations.
\end{example}

\subsection{Proof of Proposition \ref{prop: epsilon efficient}}
\begin{itemize}
    \item Proof: We show that social welfare cannot decrease by more than $v_i^T x_i(0) \cdot \mathbf{I}_{v_i^T x_i(0) < p_i(0)}$ in the $i$th iteration. Summing over all the iterations gives the result. We consider three cases.
    \begin{enumerate}
        \item EV $i$ chooses the original bundle: Social welfare remains the same. 
        \item EV $i$ chooses a different nonempty bundle: Social welfare weakly increases since both payments cancel out when comparing the utilities.
\begin{equation}
\begin{aligned}
&v_i^T x_i(1) - p_i(0) \geq v_i^T x_i(0) - p_i(0) \\
&\Rightarrow v_i^T x_i(1) \geq v_i^T x_i(0)
\end{aligned}
\end{equation}
        \item EV $i$ chooses to cancel: We must have $v_i^T x_i(0) < p_i(0)$; for otherwise, retaining the original bundle gives a higher utility. Then, social welfare decreases by $v_i^T x_i(0)$.
    \end{enumerate}

    \item A closer examination of the proof will reveal that the term $\sum_{i \in I}v_i^T x_i(0) \cdot \mathbf{I}_{v_i^T x_i(0) < p_i(0)}$ is not tight since the third case is satisfied only when $v_{i, j} < p_i(0)$ for any available bundle $j$. It is possible to formalize this more general condition in terms of the available bundle $\mathbb{T}(i)$ in each iteration. 
\end{itemize}

\subsection{Day-Ahead VCG Mechanism}
\label{app: day-ahead vcg}

We translate the real-time VCG mechanism in Section \ref{sec: pre VCG} to the day-ahead market below. In the day-ahead market, the operator runs an auction by letting EVs submit bids $b(0)$ under the following participation rules.
\begin{enumerate}[label=(\alph*)]
    \item $b_{i, j}(0) \in \mathbb{R}_+$.
    \item $b_{i, 1}(0) = 0$.
\end{enumerate}
We set the allocation and payment as follows, where $X_{-i} = \{(x_j)_{j \in I_{-i}} | x_j \in \{0, 1\}^{|\mathbb{T}|}, \mathbf{1}^Tx_j = 1,\sum_{j \in I_{-i}} \bar{x}_j \preceq N\mathbf{1}\}$. 
\begin{equation}
x(0) = x^{VCG}(b(0), X), p(0) = (p_i^{VCG}(b(0), X_{-i}))_{i \in I}.
\end{equation}

The advantage is that when there is no uncertainty, the two-stage mechanism reduces to a classical VCG mechanism, thus having desirable properties. Moreover, the day-ahead market is generally thick since we can aggregate all the bids submitted over time and announce the allocations and payments at a specified deadline, which is different from running an auction in the real-time market.

When the day-ahead market follows a VCG mechanism, it provides another natural way to set the reserve prices, whose theoretical properties are similar to that in Section \ref{sec: simple reserve}.

Compute a vector of VCG payments for each EV $i$, denoted as $(p_{i, j}^{VCG})_{j \in [|\mathbb{T}|]}$, whose $j$th element is the difference of social welfare for all other EVs when EV $i$ is absent versus when she is present but allocated $\mathbb{T}_j$. We denote by $\bar{e}_j$ the vector of allocating $\mathbb{T}_j$.
    \begin{equation}
    \begin{aligned}
        X_{-i, j} = \{&(x_j)_{j \in I_{-i}} | x_j \in \{0, 1\}^{|\mathbb{T}|}, \mathbf{1}^Tx_j = 1,  \\
        &\sum_{j \in I_{-i}} \bar{x}_j \preceq N\mathbf{1} - \bar{e}_j\}.
    \end{aligned}
\end{equation}
\begin{equation}
\begin{aligned}
&p^{VCG}_{i, j}(b(0), X_{-i}) \\
&= SW_{-i}(b_{-i}(0), X_{-i}) - SW_{-i}(b_{-i}(0), X_{-i, j}).
\end{aligned}
\end{equation}

Then, we can set 
\begin{equation}
h_i((b_{j}(1))_{j \in [i-1]}) = (p^{VCG}_{i, j}(b(0), X_{-i}))_{j \in [|\mathbb{T}|]}.
\end{equation}





%% file: final.bbl
\begin{thebibliography}{10}
\providecommand{\url}[1]{#1}
\csname url@samestyle\endcsname
\providecommand{\newblock}{\relax}
\providecommand{\bibinfo}[2]{#2}
\providecommand{\BIBentrySTDinterwordspacing}{\spaceskip=0pt\relax}
\providecommand{\BIBentryALTinterwordstretchfactor}{4}
\providecommand{\BIBentryALTinterwordspacing}{\spaceskip=\fontdimen2\font plus
\BIBentryALTinterwordstretchfactor\fontdimen3\font minus \fontdimen4\font\relax}
\providecommand{\BIBforeignlanguage}[2]{{%
\expandafter\ifx\csname l@#1\endcsname\relax
\typeout{** WARNING: IEEEtran.bst: No hyphenation pattern has been}%
\typeout{** loaded for the language `#1'. Using the pattern for}%
\typeout{** the default language instead.}%
\else
\language=\csname l@#1\endcsname
\fi
#2}}
\providecommand{\BIBdecl}{\relax}
\BIBdecl

\bibitem{7548363}
E.~Bitar and Y.~Xu, ``Deadline differentiated pricing of deferrable electric loads,'' \emph{IEEE Transactions on Smart Grid}, vol.~8, no.~1, pp. 13--25, 2017.

\bibitem{9424480}
L.~Hou, J.~Yan, C.~Wang, and L.~Ge, ``A simultaneous multi-round auction design for scheduling multiple charges of battery electric vehicles on highways,'' \emph{IEEE Transactions on Intelligent Transportation Systems}, vol.~23, no.~7, pp. 8024--8036, 2022.

\bibitem{10.5555/2031678.2031733}
E.~H. Gerding, V.~Robu, S.~Stein, D.~C. Parkes, A.~Rogers, and N.~R. Jennings, ``Online mechanism design for electric vehicle charging,'' in \emph{The 10th International Conference on Autonomous Agents and Multiagent Systems - Volume 2}, ser. AAMAS '11.\hskip 1em plus 0.5em minus 0.4em\relax Richland, SC: International Foundation for Autonomous Agents and Multiagent Systems, 2011, p. 811–818.

\bibitem{7914741}
S.~Zhao, X.~Lin, and M.~Chen, ``Robust online algorithms for peak-minimizing ev charging under multistage uncertainty,'' \emph{IEEE Transactions on Automatic Control}, vol.~62, no.~11, pp. 5739--5754, 2017.

\bibitem{9840998}
C.~Lu, J.~Wu, J.~Cui, Y.~Xu, C.~Wu, and M.~C. Gonzalez, ``Deadline differentiated dynamic ev charging price menu design,'' \emph{IEEE Transactions on Smart Grid}, vol.~14, no.~1, pp. 502--516, 2023.

\bibitem{LEE2020106694}
\BIBentryALTinterwordspacing
Z.~J. Lee, J.~Z. Pang, and S.~H. Low, ``Pricing ev charging service with demand charge,'' \emph{Electric Power Systems Research}, vol. 189, p. 106694, 2020. [Online]. Available: \url{https://www.sciencedirect.com/science/article/pii/S0378779620304971}
\BIBentrySTDinterwordspacing

\bibitem{9875037}
Z.~Ding, Y.~Zhang, W.~Tan, X.~Pan, and H.~Tang, ``Pricing based charging navigation scheme for highway transportation to enhance renewable generation integration,'' \emph{IEEE Transactions on Industry Applications}, vol.~59, no.~1, pp. 108--117, 2023.

\bibitem{10542458}
M.~H. Abbasi, Z.~Arjmandzadeh, J.~Zhang, V.~Krovi, B.~Xu, and D.~K. Mishra, ``A coupled game theory and lyapunov optimization approach to electric vehicle charging at fast charging stations,'' \emph{IEEE Transactions on Vehicular Technology}, pp. 1--11, 2024.

\bibitem{1eba3f1b-3634-3b34-9a46-c86b8d069d2d}
\BIBentryALTinterwordspacing
P.~Cramton, ``Electricity market design,'' \emph{Oxford Review of Economic Policy}, vol.~33, no.~4, pp. pp. 589--612, 2017. [Online]. Available: \url{https://www.jstor.org/stable/48539475}
\BIBentrySTDinterwordspacing

\bibitem{8315146}
Z.~Liu, Q.~Wu, K.~Ma, M.~Shahidehpour, Y.~Xue, and S.~Huang, ``Two-stage optimal scheduling of electric vehicle charging based on transactive control,'' \emph{IEEE Transactions on Smart Grid}, vol.~10, no.~3, pp. 2948--2958, 2019.

\bibitem{6740918}
L.~Yang, J.~Zhang, and H.~V. Poor, ``Risk-aware day-ahead scheduling and real-time dispatch for electric vehicle charging,'' \emph{IEEE Transactions on Smart Grid}, vol.~5, no.~2, pp. 693--702, 2014.

\bibitem{7394192}
R.~Wang, P.~Wang, and G.~Xiao, ``Two-stage mechanism for massive electric vehicle charging involving renewable energy,'' \emph{IEEE Transactions on Vehicular Technology}, vol.~65, no.~6, pp. 4159--4171, 2016.

\bibitem{WU201755}
\BIBentryALTinterwordspacing
F.~Wu and R.~Sioshansi, ``A two-stage stochastic optimization model for scheduling electric vehicle charging loads to relieve distribution-system constraints,'' \emph{Transportation Research Part B: Methodological}, vol. 102, pp. 55--82, 2017. [Online]. Available: \url{https://www.sciencedirect.com/science/article/pii/S0191261516305343}
\BIBentrySTDinterwordspacing

\bibitem{7360236}
L.~Zhang and Y.~Li, ``Optimal management for parking-lot electric vehicle charging by two-stage approximate dynamic programming,'' \emph{IEEE Transactions on Smart Grid}, vol.~8, no.~4, pp. 1722--1730, 2017.

\bibitem{8293857}
Y.~Wang and J.~S. Thompson, ``Two-stage admission and scheduling mechanism for electric vehicle charging,'' \emph{IEEE Transactions on Smart Grid}, vol.~10, no.~3, pp. 2650--2660, 2019.

\bibitem{7833208}
D.~Wu, H.~Zeng, C.~Lu, and B.~Boulet, ``Two-stage energy management for office buildings with workplace ev charging and renewable energy,'' \emph{IEEE Transactions on Transportation Electrification}, vol.~3, no.~1, pp. 225--237, 2017.

\bibitem{8960365}
X.~Wang, C.~Sun, R.~Wang, and T.~Wei, ``Two-stage optimal scheduling strategy for large-scale electric vehicles,'' \emph{IEEE Access}, vol.~8, pp. 13\,821--13\,832, 2020.

\bibitem{TAN2022107359}
\BIBentryALTinterwordspacing
B.~Tan, H.~Chen, X.~Zheng, and J.~Huang, ``Two-stage robust optimization dispatch for multiple microgrids with electric vehicle loads based on a novel data-driven uncertainty set,'' \emph{International Journal of Electrical Power \& Energy Systems}, vol. 134, p. 107359, 2022. [Online]. Available: \url{https://www.sciencedirect.com/science/article/pii/S0142061521005986}
\BIBentrySTDinterwordspacing

\bibitem{LIN2023104715}
\BIBentryALTinterwordspacing
H.~Lin, J.~Dang, H.~Zheng, L.~Yao, Q.~Yan, S.~Yang, H.~Guo, and A.~Anvari-Moghaddam, ``Two-stage electric vehicle charging optimization model considering dynamic virtual price-based demand response and a hierarchical non-cooperative game,'' \emph{Sustainable Cities and Society}, vol.~97, p. 104715, 2023. [Online]. Available: \url{https://www.sciencedirect.com/science/article/pii/S2210670723003268}
\BIBentrySTDinterwordspacing

\bibitem{6808513}
M.~Moeini-Aghtaie, A.~Abbaspour, and M.~Fotuhi-Firuzabad, ``Online multicriteria framework for charging management of phevs,'' \emph{IEEE Transactions on Vehicular Technology}, vol.~63, no.~7, pp. 3028--3037, 2014.

\bibitem{10378662}
Y.~Kabiri-Renani, A.~Arjomandi-Nezhad, M.~Fotuhi-Firuzabad, and M.~Shahidehpour, ``Transactive-based day-ahead electric vehicles charging scheduling,'' \emph{IEEE Transactions on Transportation Electrification}, vol.~10, no.~4, pp. 8235--8245, 2024.

\bibitem{7811294}
T.~Conway, ``On the effects of a routing and reservation system on the electric vehicle public charging network,'' \emph{IEEE Transactions on Intelligent Transportation Systems}, vol.~18, no.~9, pp. 2311--2318, 2017.

\bibitem{smartcities3040067}
\BIBentryALTinterwordspacing
R.~Basmadjian, B.~Kirpes, J.~Mrkos, and M.~Cuchý, ``A reference architecture for interoperable reservation systems in electric vehicle charging,'' \emph{Smart Cities}, vol.~3, no.~4, pp. 1405--1427, 2020. [Online]. Available: \url{https://www.mdpi.com/2624-6511/3/4/67}
\BIBentrySTDinterwordspacing

\bibitem{en13123263}
\BIBentryALTinterwordspacing
S.~Orcioni and M.~Conti, ``Ev smart charging with advance reservation extension to the ocpp standard,'' \emph{Energies}, vol.~13, no.~12, 2020. [Online]. Available: \url{https://www.mdpi.com/1996-1073/13/12/3263}
\BIBentrySTDinterwordspacing

\bibitem{s22082834}
\BIBentryALTinterwordspacing
R.~Flocea, A.~Hîncu, A.~Robu, S.~Senocico, A.~Traciu, B.~M. Remus, M.~S. Răboacă, and C.~Filote, ``Electric vehicle smart charging reservation algorithm,'' \emph{Sensors}, vol.~22, no.~8, 2022. [Online]. Available: \url{https://www.mdpi.com/1424-8220/22/8/2834}
\BIBentrySTDinterwordspacing

\bibitem{8850812}
Y.~Cao, S.~Liu, Z.~He, X.~Dai, X.~Xie, R.~Wang, and S.~Yu, ``Electric vehicle charging reservation under preemptive service,'' in \emph{2019 1st International Conference on Industrial Artificial Intelligence (IAI)}, 2019, pp. 1--6.

\bibitem{8734897}
Y.~Cao, T.~Jiang, O.~Kaiwartya, H.~Sun, H.~Zhou, and R.~Wang, ``Toward pre-empted ev charging recommendation through v2v-based reservation system,'' \emph{IEEE Transactions on Systems, Man, and Cybernetics: Systems}, vol.~51, no.~5, pp. 3026--3039, 2021.

\bibitem{LIU2021103150}
\BIBentryALTinterwordspacing
S.~Liu, X.~Xia, Y.~Cao, Q.~Ni, X.~Zhang, and L.~Xu, ``Reservation-based ev charging recommendation concerning charging urgency policy,'' \emph{Sustainable Cities and Society}, vol.~74, p. 103150, 2021. [Online]. Available: \url{https://www.sciencedirect.com/science/article/pii/S2210670721004327}
\BIBentrySTDinterwordspacing

\bibitem{9068444}
X.~Zhang, Y.~Cao, L.~Peng, J.~Li, N.~Ahmad, and S.~Yu, ``Mobile charging as a service: A reservation-based approach,'' \emph{IEEE Transactions on Automation Science and Engineering}, vol.~17, no.~4, pp. 1976--1988, 2020.

\bibitem{7387372}
Y.~Cao, N.~Wang, Y.~J. Kim, and C.~Ge, ``A reservation based charging management for on-the-move ev under mobility uncertainty,'' in \emph{2015 IEEE Online Conference on Green Communications (OnlineGreenComm)}, 2015, pp. 11--16.

\bibitem{7756376}
Y.~Cao, T.~Wang, O.~Kaiwartya, G.~Min, N.~Ahmad, and A.~H. Abdullah, ``An ev charging management system concerning drivers’ trip duration and mobility uncertainty,'' \emph{IEEE Transactions on Systems, Man, and Cybernetics: Systems}, vol.~48, no.~4, pp. 596--607, 2018.

\bibitem{BERNAL2020100388}
\BIBentryALTinterwordspacing
R.~Bernal, D.~Olivares, M.~Negrete-Pincetic, and Álvaro Lorca, ``Management of ev charging stations under advance reservations schemes in electricity markets,'' \emph{Sustainable Energy, Grids and Networks}, vol.~24, p. 100388, 2020. [Online]. Available: \url{https://www.sciencedirect.com/science/article/pii/S2352467720303192}
\BIBentrySTDinterwordspacing

\bibitem{Nisan_Roughgarden_Tardos_Vazirani_2007}
N.~Nisan, T.~Roughgarden, E.~Tardos, and V.~V. Vazirani, \emph{Algorithmic Game Theory}.\hskip 1em plus 0.5em minus 0.4em\relax Cambridge University Press, 2007.

\bibitem{9381520}
T.~Zeng, S.~Bae, B.~Travacca, and S.~Moura, ``Inducing human behavior to maximize operation performance at pev charging station,'' \emph{IEEE Transactions on Smart Grid}, vol.~12, no.~4, pp. 3353--3363, 2021.

\bibitem{MYERSON1983265}
\BIBentryALTinterwordspacing
R.~B. Myerson and M.~A. Satterthwaite, ``Efficient mechanisms for bilateral trading,'' \emph{Journal of Economic Theory}, vol.~29, no.~2, pp. 265--281, 1983. [Online]. Available: \url{https://www.sciencedirect.com/science/article/pii/0022053183900480}
\BIBentrySTDinterwordspacing

\bibitem{10.1257/000282803322157061}
\BIBentryALTinterwordspacing
A.~Abdulkadiroğlu and T.~Sönmez, ``School choice: A mechanism design approach,'' \emph{American Economic Review}, vol.~93, no.~3, p. 729–747, June 2003. [Online]. Available: \url{https://www.aeaweb.org/articles?id=10.1257/000282803322157061}
\BIBentrySTDinterwordspacing

\bibitem{10.1093/oxrep/grx063}
\BIBentryALTinterwordspacing
S.~D. Kominers, A.~Teytelboym, and V.~P. Crawford, ``{An invitation to market design},'' \emph{Oxford Review of Economic Policy}, vol.~33, no.~4, pp. 541--571, 11 2017. [Online]. Available: \url{https://doi.org/10.1093/oxrep/grx063}
\BIBentrySTDinterwordspacing

\bibitem{https://doi.org/10.1111/j.1468-0262.2006.00695.x}
\BIBentryALTinterwordspacing
S.~Bikhchandani, S.~Chatterji, R.~Lavi, A.~Mu'alem, N.~Nisan, and A.~Sen, ``Weak monotonicity characterizes deterministic dominant-strategy implementation,'' \emph{Econometrica}, vol.~74, no.~4, pp. 1109--1132, 2006. [Online]. Available: \url{https://onlinelibrary.wiley.com/doi/abs/10.1111/j.1468-0262.2006.00695.x}
\BIBentrySTDinterwordspacing

\end{thebibliography}
